# Structural Twinning-induced Insulating Phase in CrN (111) Films

Qiao Jin,[1,2,#] Zhiwen Wang,[3,#] Qinghua Zhang,[1] Jiali Zhao,[1] Hu Cheng,[4] Shan Lin,[1,2] Shengru Chen, [1,2] Shuang Chen,[5] Haizhong Guo,[5] Meng He,[1] Chen Ge,[1] Can Wang,[1,2,6] Jia-Ou Wang,[7] Lin Gu,[1,2,6] Shanmin Wang,[4] Hongxin Yang,[3,*] Kui-juan Jin,[1,2,6,*] and Er-Jia Guo[1,6,8,*]

[1] *Beijing National Laboratory for Condensed Matter Physics and Institute of Physics, Chinese Academy of Sciences, Beijing 100190, China*

[2] *School of Physical Sciences, University of Chinese Academy of Sciences, Beijing 100190, China*

[3] *Ningbo Institute of Materials Technology & Engineering, Chinese Academy of Sciences, Ningbo 315201, China*

[4] *Department of Physics, Southern University of Science and Technology, Shenzhen, Guangdong 518055, China*

[5] *School of Physics and Microelectronics, Zhengzhou University, Zhengzhou 450001, China*

[6] *Songshan Lake Materials Laboratory, Dongguan, Guangdong 523808, China*

[7] *Institute of High Energy Physics, Chinese Academy of Sciences, Beijing 100049, China*

[8] *Center of Materials Science and Optoelectronics Engineering, University of Chinese Academy of Sciences, Beijing 100049, China*

*Correspondence and requests for materials should be addressed to H. X. Y., K. J., and E. J. G. (emails: hongxin.yang@nimte.ac.cn, kjjin@iphy.ac.cn and ejguo@iphy.ac.cn)

**Abstract**

Electronic states of a correlated material can be effectively modified by structural variations delivered from a single-crystal substrate. In this letter, we show that the CrN films grown on MgO (001) substrates have a (001) orientation, whereas the CrN films on α-Al$_2$O$_3$ (0001) substrates are oriented along (111) direction parallel to the surface normal. Transport properties of CrN films are remarkably different depending on crystallographic orientations. The critical thickness for the metal-insulator transition (MIT) in CrN 111 films is significantly larger than that of CrN 001 films. In contrast to CrN 001 films without apparent defects, scanning transmission electron microscopy results reveal that CrN 111 films exhibit strain-induced structural defects, e. g. the periodic horizontal twinning domains, resulting in an increased electron scattering facilitating an insulating state. Understanding the key parameters that determine the electronic properties of ultrathin conductive layers is highly desirable for future technological applications.

**Main text**

Chromium nitride (CrN) with its impressive mechanical and thermal properties is a promising material for future technological applications in cutting, coating, and thermoelectric industrial products. [1, 2] In addition, CrN is an antiferromagnet below the Néel temperature ($T_N$) of ~280 K. It produces zero magnetization and zero parasitic stray field and is robust against magnetic perturbations. These characters make CrN is ideal for secure data storage and memory devices. [3-5] Compared to other well-established metallic antiferromagnets, such as MnPt, IrMn, $Mn_2Au$, [6-8] CrN does not consist of noble elements. In contrast, CrN is cheap, chemically stable, and corrosion resistive, which are the central requirements for developing functional devices. Therefore, understanding of thickness, strain, and orientation effects on the structural and electronic properties of CrN films is indispensable for their practical applications.

Strain engineering on strongly correlated materials has provided an effective control over the materials structure and stoichiometry, which leads to a wide spectrum of intriguing phenomena and functionalities of thin films. [9-12] High-quality single crystalline substrates stabilize multiple phases by statically fixing lattice distortions in epitaxial thin films and also select preferential crystalline orientations owing to the appropriate symmetry and minimized misfit strain. CrN has a cubic rock-salt structure (Figure 1a) and its lattice constant is 4.15 Å at room temperature. [1,2,13-17] The CrN films grown on MgO (001) substrates follow the substrate's orientation. The lattice mismatch between CrN and MgO ($a$ = 4.21 Å) is 1.4%, resulting in a moderate tensile strain.[18] On the other hand, when viewed along the [111] direction, CrN consists of alternating planes of $Cr^{3+}$ and $N^{3-}$ ions with a hexagonal structure (Figure 1b), which is similar to that of the (0001) surface of sapphire (α−$Al_2O_3$, *R−3c*). [19, 20] The atomic distance between two nearby $Cr^{3+}$ ions is 2.93 Å, which is close to the atomic distance between two $Al^{3+}$ ions (2.77 Å) in α−$Al_2O_3$ (Figure 1c).[21] The (111)-oriented CrN film grown on α−$Al_2O_3$ will suffer an extremely large compressive strain up to −5.4%. Therefore, we are motivated to fabricate CrN films on both MgO and α−$Al_2O_3$ substrates. Firstly, it is challenging to fabricate high-quality single crystalline CrN films under such a huge misfit strain. The mechanism of strain relaxation in (111)-oriented CrN films is unexplored. Secondly, the growth of CrN films with identical thickness allows us to directly compare their physical properties under both tensile and compressive strains. Lastly, the thickness dependent electronic state and local structure of (111)-oriented CrN films have never been reported so far. In this paper, we investigate the thickness-dependent electronic state in CrN films with (001) and (111) orientations. The critical thickness for the metal-to-insulator transition (MIT) of CrN 111 films is significantly larger than that of CrN 001 films. Microstructural characterizations reveal that the CrN 111 films possess structural twining domains along the growth direction, which results in enhanced electron scattering and dramatic change in critical behaviors.

CrN films were simultaneously grown on (001)−oriented MgO and (0001)−oriented α-$Al_2O_3$ substrates by active-nitrogen-assisted pulsed laser deposition. A stoichiometric ceramic CrN was used as a target. It was synthesized using a high-pressure reaction route from a mixed $CrCl_3$ and $NaNH_2$ powder. [17] The heating process was performed at 5 GPa and 1200 °C for 5 mins. Then the powder was sintered as a 1-inch-diameter target at 5 GPa and 1100 °C for 50 mins. Same process was used for synthesizing the stoichiometric single-crystalline nitrides as described in our earlier work. [17] To obtain highly crystalline CrN thin films, the growth conditions were optimized by systematically varying the temperature, laser fluence, and film–target distance. The best quality CrN films were fabricated under the substrate-target distance of 7 cm, the substrate temperature of 600°C, laser fluence of ~1.3 J/cm$^2$, and low base pressure of $1 \times 10^{-8}$ Torr. During the film deposition, the base pressure increases to $5 \times 10^{-7}$ Torr due to the substrate's heating. After the growth, the CrN films were cooled down in vacuum. The film thickness was controlled by counting the number of laser pulses. X-ray reflectivity and diffraction measurements were carried out on a Bruker D8 diffractometer equipped with monochromator using Cu target ($\lambda = 1.54$ Å) (see Fig. S1 in the Supplemental Material [22]). Figure 1d shows the wide-range X-ray diffraction (XRD) $\theta-2\theta$ scans for CrN films with the thickness of ~ 20 nm grown on MgO and α−Al$_2$O$_3$ substrates. No impurity phase was observed in both CrN films. The CrN films grown on MgO (001) substrates are (001) oriented, whereas the CrN films grown on α-Al$_2$O$_3$ substrates have (111) orientation along the surface normal. The insets in Figure 1d show the high-magnified XRD curves around the 001 and 111 peaks of CrN films. The clear Laue fringes indicate the high crystalline quality of as-grown CrN films. Omega scans were used to determine the full width at half maximum (FWHM) values for the 001 and 111 peaks of CrN films. The FWHM values for CrN films are ~0.04°, which are comparable to those of the single-crystalline substrates (~0.02°). The out-of-plane lattice constants ($c$) of CrN films are calculated precisely by applying Nelson Riley function (see Fig. S2 in the Supplemental Material [22]). $c$ of 20-nm-thick (00$L$)-oriented CrN film and ($LLL$)-oriented CrN film are 4.117 Å and 2.412 Å, respectively. As increasing the film thickness from 2.4 to 120 nm, the out-of-plane lattice constants of CrN 111 films reduce towards its bulk value ($c_{bulk}$ ~ 2.405 Å), indicating the gradual release of compressive strain. In addition, phi-scans were performed to investigate the in-plane epitaxial relationship between CrN films and substrates. CrN 001 film exhibits a cubic symmetry, whereas the CrN 111 film yields a hexagonal symmetry. Both CrN films show a good in-plane epitaxial growth on the substrates.

It is known that chromium nitrides have parasitic phases, e. g. $Cr_2N$,[23-25] or $CrN_{1-x}$ phase induced by nitrogen vacancies under reduced deposition conditions. To obtained direct information about the chemical composition of CrN films, we carried out secondary-ion mass spectrometry (SIMS) measurements on a 50 nm-thick CrN 111 film (see Fig. S3 in the Supplemental Material [22]). The $Ce^{3+}$-ion beam was rastered over a region of about 250 × 250 μm$^2$. The concentration of $N^{3-}$ ions was pre-calibrated using CrN ceramic samples. The molar ratio between $Cr^{3+}$ and $N^{3-}$ ions is close to 1:1. X-ray absorption spectroscopy (XAS) measurements were performed at the N *K*-edges and Cr *L*-edges for CrN films with a thickness of 50 nm at room temperature. The instrumental resolution of our XAS measurements is 200 meV. The reference spectra from stoichiometric CrN films were shown for direct comparison (Figure 1e). XAS results reveal that the valence state of Cr ions is +3, and the upper limit of $Cr^{2+}$ ions in the CrN films is 2% (see Fig. S4 in the Supplemental Material [22]). Therefore, SIMS and XAS indicate the chemical formular of as-grown films to be CrN and the concentration of nitrogen vacancies is beyond the detection limit of those techniques.

The transport properties of CrN films were measured using the standard van der Pauw method. [26] Figure 2a shows the temperature-dependent resistivity ($\rho$) of CrN 001 and CrN 111 films with the thickness of ~20 nm. The CrN 001 film exhibits a metallic behavior at all temperatures, and $\rho$ undergoes a clear transition at ~150 K, which corresponds to the $T_N$ of CrN films. Of note, the $T_N$ of an ultrathin CrN film deviates from its bulk value (~270 K). [13-16] This behavior can be attributed to the finite-size effect in thin films, which is similar to the behavior of other antiferromagnetic thin films such as CoO [27, 28] and Mn$_3$Ir [29]. In sharp contrast to the metallic phases in CrN 001 films, the 20-nm-thick CrN 111 film exhibits semiconducting behavior (d$\rho$/d$T$ < 0) with a decrease in temperature. We estimated that the thermal activation energy of charge carriers in 20-nm-thick CrN 111 is ~39.15 meV by linearly fitting the $ln(\rho)-(T)^{-1}$ curve (inset of Figure 2a). Figure 2b shows the $\rho$-$T$ curves of CrN 111 films with various film thickness (*t*) ranging from 2.4 to 120 nm. The semiconducting behavior is noticeable in all CrN111 films when *t* is below 50 nm. With an increase in *t*, $\rho_{300K}$ monotonically decreases. For a 50-nm-thick CrN 111 film, a low-temperature upturn is observed at ~50 K. The 120-nm-thick CrN 111 film exhibits a metallic transport character at all temperatures. Similar to a previous study, we do not observe resistivity discontinuity with a decrease in temperature. [18] Therefore, CrN 111 films do not show an anomaly in metallic conductivity around $T_N$ because the (0001) surface of α-Al$_2$O$_3$ prevents the CrN 111 film from distorting its atomic structure. The resistivity of 2.4-nm-thick CrN 111 film increases by five orders of magnitude when *T* decreases from 300 to 10 K. Thus, MIT occurs in the CrN 111 films with a decrease in film thickness. Our previous study demonstrates that the $T_N$ increases with film thickness and the critical thickness for MIT in CrN 001 films is approximately 12 nm [30] as summarized in Figures 2c and 2d. The critical thickness for MIT in CrN 111 films increases up to 50 nm (Figure 2e), which is much larger than the electrical dead layer thickness (~ 5 u.c., i. e. 2 nm) of most metallic thin films.[31-33] These results are consistent with our earlier work, which showed that large compressive strain preferentially promoted an insulating state in the CrN films owing to the strong modification of orbital splitting.[30]

Microstructural characterizations on CrN films with different orientations were conducted using scanning transmission electron microscopy (STEM). Figures 3a and 3b show the high-angle annular dark field (HAADF) STEM images of a 20-nm-thick CrN 001 and a 50-nm-thick CrN 111 films, respectively. Of note, the HAADF−STEM image provides scattering intensities that are approximately proportional to the square of atomic number. Thus, the bright features in HAADF images indicate the positions of the Cr atom column. The light element N cannot be observed in HAADF images. In Figures 3c-3f, high-magnified HAADF−STEM results confirm that all CrN films are epitaxially grown on the substrates with atomically sharp interfaces and CrN films exhibit high crystalline quality. For CrN 001 films, the high-magnified image and line profile of Cr intensity, shown in Figure 3g, exhibit the well-aligned atomic structure without any detectable defects and misfit dislocations. Figure 3h indicates that the distance between two neighboring Cr planes is ~2.41 Å, which is in agreement with XRD results. In addition, CrN 111 films possess the layered structure by forming horizontal structural domains with a period of 3 u.c. (i. e. every three-unit-cells would appear a structural defect). Thus, we estimated that the density of these structural defects in the CrN 111 films consist of about 33%. We determined that each domain misaligned along the in-plane direction by ~1.5 Å. This distance equals approximately 1/2 of the distance between two $Cr^{3+}$ ions. The domain boundaries are clearly visible and highlighted by yellow dashed lines in Figure 3e. The misaligned domains are absence along the in-plane direction. The crystal structures of Cr and N atoms are sketched in both Figures 3c and 3e to clarify the atomic arrangement. The annular bright field (ABF) STEM image of a representative region in the CrN 111 films is shown in Figure S5 of Supplementary Materials.[22] The light element N can be distinguished in the ABF−STEM image, which is in excellent agreement with our atomic positions. These structural twins were frequently observed in delafossite oxide thin films, such as $PdCoO_2$ and $PdCrO_2$, grown on α−$Al_2O_3$ substrates. [34, 35]

The origin of twining domains in the CrN 111 films can be attributed to the extremely large compressive misfit strain (−5.4%) between the CrN films and α-$Al_2O_3$ substrates. The formation of periodic structural twins indicates that their formation energies are relatively small and equivalent, which means that they epitaxially match substrate's crystalline symmetry and hexagonal lattice structure. The unique domain structure allows compressive strain to persist in a relatively thick CrN film. Therefore, the critical thickness for MIT in CrN 111 films is considerably larger than that in CrN 001 films. The existence of domain boundaries enhances the probability for electron scattering; thus, CrN 111 films are insulating with a relatively large film thickness.

To understand the influence of defects on the electronic states of CrN films, we performed the first-principles calculations within the framework of density-functional theory (DFT) as implemented in the Vienna *ab initio* simulation package. [36, 37] Here, we adopt the value of $U − J = 3$ eV [30, 38] for the DFT + $U$ calculations. We take both the (001)− and (111)−oriented CrN films with the same thickness of 10 u.c. into consideration. We had tested four different magnetic orders with different spin polarizations (Table I and Figure S6 of Supplementary Materials, Ref. [22]). Similar to the previous theoretical works, [39] the most stable magnetic structure is C-type AFM with relative energy equals to zero. The AFM state is also consistent with our recent experimental observations. [30] In the DFT calculations, we constrained the in-plane lattice constant of CrN to that of substrates. CrN 001 film is under a tensile strain of 1.4%, whereas the CrN 111 film is compressively strained (-5.4%). Figure 4 shows the band structures of the CrN films with (001)− and (111)−orientations. For the CrN 001 film, a hole pocket formed at the *M* point, and an electron pocket formed at the *Γ* point. Moreover, the clear band crossings at the Fermi level are observed in the CrN 001 films. However, the CrN 111 film shows a relatively flat band curvature around the Fermi level, and a band gap exists. We calculate the band gaps in both CrN films by subtracting the energy difference between the values at the bottom of the conduction band and at the top of the valence band. The bandgap in the CrN 001 is −13 meV [in Figure 4a], which suggests the existence of metallic state. However, CrN 111 with the same thickness has the bandgap of 66 meV, as shown in Figure 4b, which demonstrates that the CrN 111 film is a semiconductor. With an increase in the thickness of CrN 001 films, a previous study [30] determined that starting with a critical thickness of 10 u.c., the gap closed, and conducting states appeared. However, the band structure of CrN 111 thin films shows an insulating state at the same thickness. Therefore, CrN 111 films are more insulating than CrN 001 films. Of note, DFT calculations help to quantitatively understand our experimental observations. The tendency of bandgap to open by altering the crystalline orientation (i. e. the epitaxial strain) should not change with film thickness.

In summary, we report the orientation-dependent electronic state in high-quality single crystalline CrN films. We determine that both (001)− and (111)−oriented CrN films undergo a MIT with a decrease in the film thickness. For the thickest CrN films, the residual-resistivity-ratio (RRR) of CrN 001 film is 2.28, whereas the RRR of CrN 111 film is only 1.3. We believe that the extremely large compressive strain between CrN and $Al_2O_3$ results in the periodic structural twining domains. These structural defects are responsible for the reduced RRR value in CrN 111 films. Our results are supported by the theoretical calculations, confirming that CrN 111 films are more insulating than CrN 001 films for the same film thickness. Our results demonstrate that crystalline orientation is essential for tailoring functional materials to design magnetic and electrical materials with superior physical properties.

**Acknowledgements**

This work was supported by the National Key Basic Research Program of China (Grant Nos. 2020YFA0309100 and 2019YFA0308500), the National Natural Science Foundation of China (Grant Nos. 11974390, 52025025, and 52072400), the Beijing Nova Program of Science and Technology (Grant No. Z191100001119112), the Beijing Natural Science Foundation (Grant No. 2202060), and the Strategic Priority Research Program (B) of the Chinese Academy of Sciences (Grant No. XDB33030200). The XAS experiments at the beam line 4B9B of the Beijing Synchrotron Radiation Facility (BSRF) of the Institute of High Energy Physics, Chinese Academy of Sciences were conducted via a user proposal.

**Data availability**

The data that supports the findings of this study are openly available in the Wiley website [Q. Jin, H. Cheng, Z. Wang, Q. Zhang, S. Lin, M. A. Roldan, J. Zhao, J. -O. Wang, S. Chen, M. He, C. Ge, C. Wang, H. -B. Lu, H. Guo, L. Gu, X. Tong, T. Zhu, S. Wang, H. Yang, K. -j. Jin, E. -J. Guo, Strain-mediated high conductivity in ultrathin antiferromagnetic metallic nitrides, Advanced Materials 33, 2005920 (2021)], and we properly cite this publication in the reference number [30].

**Figure and figure captions**

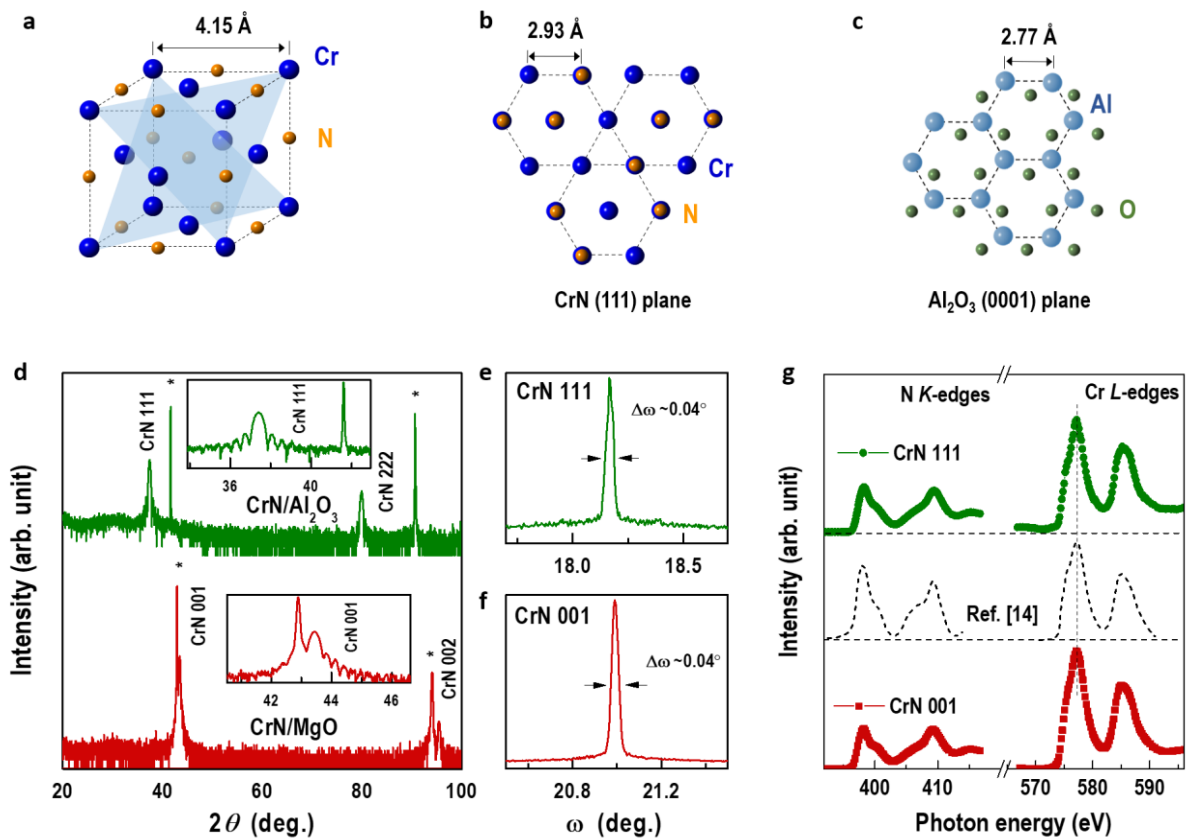

**Figure 1. Structural and electronic state characterization for (001)- and (111)-oriented CrN films.** (a) Crystal structure of CrN (space group F$m$-$3m$). Blue and orange spheres represent the Cr and N atoms, respectively. (b) and (c) Planar views of the CrN and Al$_2$O$_3$ crystal structure for (111)- and (0001)-orientations, respectively. (d) XRD $\theta$-$2\theta$ scans of 20-nm-thick CrN films grown on Al$_2$O$_3$ (0001) and MgO (001) substrates. The diffraction peaks from the substrates are indicated by asterisk (*). The curves are offset for clarity. Insets show the high-magnified scans around 111 and 001 peaks of the CrN film grown on Al$_2$O$_3$ and MgO substrates. The corresponding rocking curves for (e) CrN 111 and (f) CrN 001 peaks, respectively. The FWHMs of both rocking curves are ~0.04°. (g) XAS at N $K$- and Cr $L$-edges for the (001)- and (111)-oriented 12-nm-thick CrN films. The reference data from Ref. [14] is shown for direct comparison.

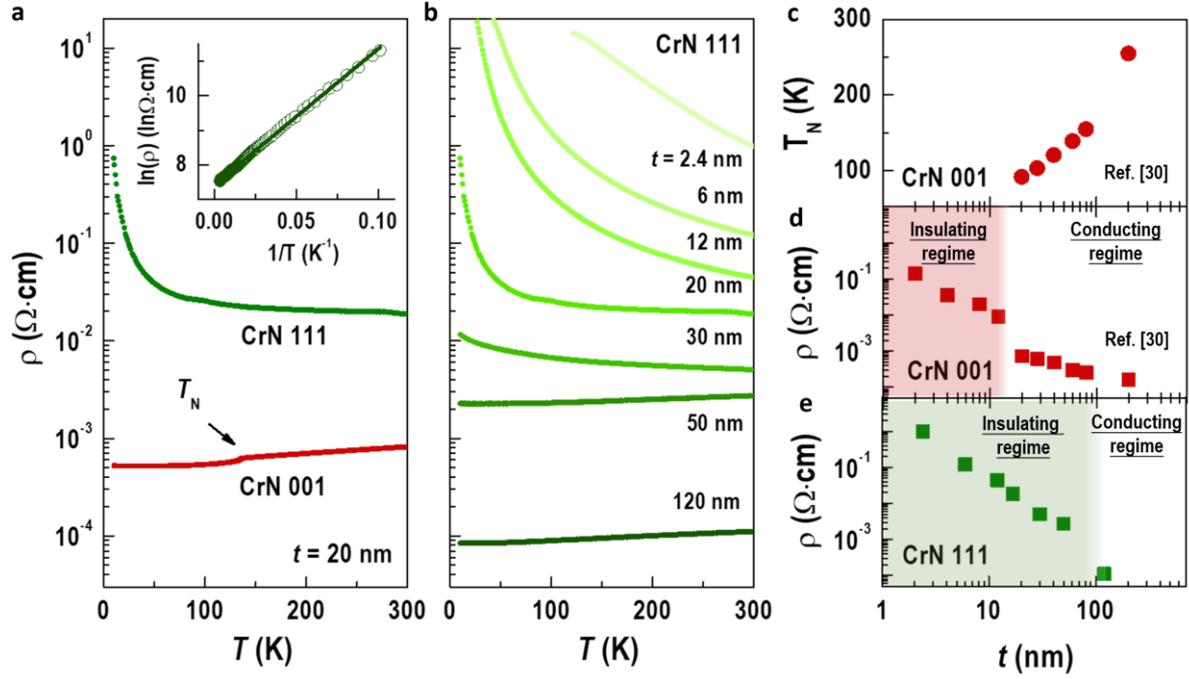

**Figure 2. Transport properties of CrN films.** (a) $\rho$-T curves of the (001)- and (111)-oriented CrN films with the thickness of 20 nm. Inset: the linear fit of $ln(\rho)-(T)^{-1}$ curve of a 20-nm-thick CrN 111 film. (b) $\rho$-T curves of the CrN 111 films with the thickness ranging from 2.4 to 120 nm. (c) $T_N$ and (d) $\rho$ of (001)-oriented CrN film as a function of film thickness. (e) $\rho$ of (111)-oriented CrN film as a function of film thickness. Previous studies have shown that the critical thickness for metal-to-insulator transition (MIT) in the CrN 001 films is approximately 12 nm,[30] whereas the critical thickness of MIT in the CrN 111 films increases to 50 nm.

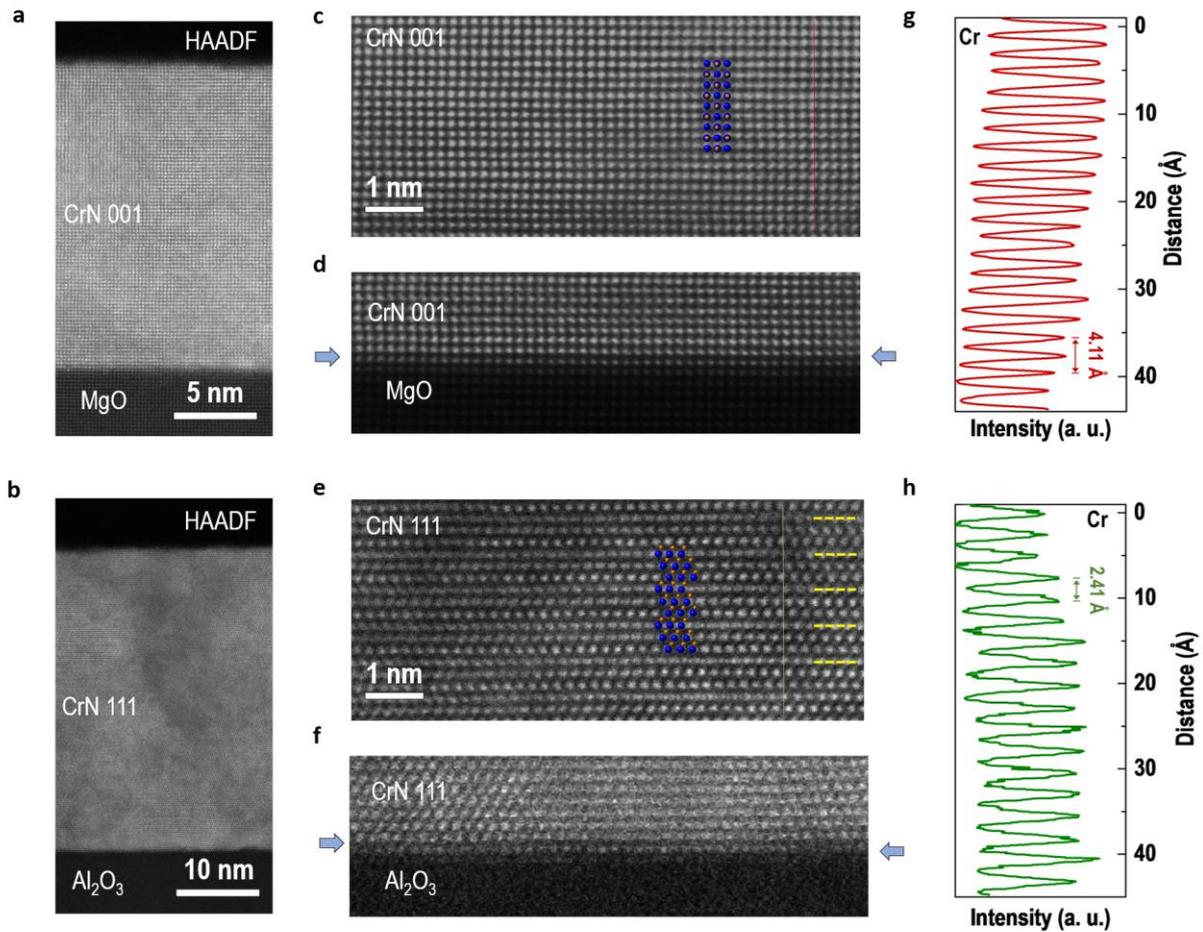

**Figure 3. Microstructural characterizations of CrN films.** High-angle annular dark field (HAADF) scanning transmission electron microscopy (STEM) images of (a) (001)- and (b) (111)-oriented CrN films. The HAADF-STEM images of CrN 001 and CrN 111 films are acquired along the [100] and [−1100] zone axis, respectively. (c) and (d) High-magnified STEM images of CrN 001 film bulk and CrN-MgO interface, respectively. The layer-resolved integrated red line profile is plotted in (g). (e) and (f) High-magnified STEM images of CrN 111 film bulk and CrN-Al$_2$O$_3$ interface, respectively. The yellow dashed lines indicate the positions of structural twins. The layer-resolved integrated green line profile is plotted in (h).

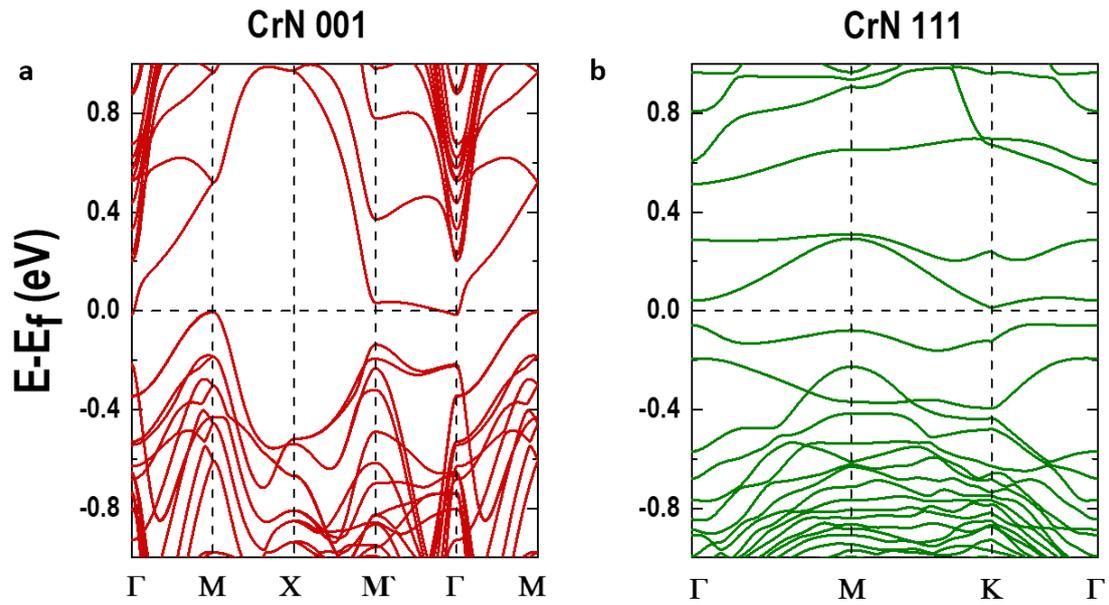

**Figure 4. Orientation-dependent band structure of CrN films.** (a) and (b) Band structures of CrN films with the fixed thickness of 10 u.c. The bandgap in CrN 001 films is −13 meV, which suggests a metallic state. In contrast, the bandgap in CrN 111 films with an identical thickness is 66 meV, which demonstrates an insulating state.